\documentclass{emulateapj}
\usepackage{graphicx}

\usepackage{verbatim}
\usepackage{amsmath, amsthm, amssymb}

\def\ga{\mathrel{\hbox{\rlap{\hbox{\lower4pt\hbox{$\sim$}}}\hbox{$>$}}}}
\def\la{\mathrel{\hbox{\rlap{\hbox{\lower4pt\hbox{$\sim$}}}\hbox{$<$}}}}

\shorttitle{Pressure Equilibrium HI Clumps in Virgo}

\shortauthors{Burkhart \& Loeb}
\begin{document}

\title{Predicted Sizes of Pressure-Supported HI Clouds in the Outskirts of the Virgo Cluster}

\author{Blakesley Burkhart\altaffilmark{1} \& Abraham Loeb\altaffilmark{1}, }
\altaffiltext{1}{Harvard-Smithsonian Center for Astrophysics, 60 Garden st. Cambridge, Ma, USA}

\begin{abstract}
Using data from the ALFALFA Arecibo HI survey of galaxies and the Virgo cluster X-ray pressure profiles from XMM-Newton,
 we investigate the possibility that starless dark HI clumps, also known as  ``dark galaxies",  are supported
by external pressure in the surrounding  intercluster medium.
We find that the starless HI clump masses, velocity dispersions and positions  allow these clumps to be
 in pressure equilibrium with the X-ray gas near the virial radius of the Virgo cluster.
We predict the sizes of these clumps  to range from 1-10 kpc, in agreement with the range of sizes found for spatially resolved HI starless clumps outside of Virgo.
Based on the predicted HI surface density of the Virgo sources, as well as a sample of other similar resolved ALFALFA HI dark clumps with follow up optical/radio observations,
we predict that most of the HI dark clumps are on the cusp of forming stars.
These HI sources therefore mark the transition between starless HI clouds and dwarf galaxies with stars. 
\end{abstract}


\section{Introduction}
\label{intro}

Galaxy clusters are the most massive virialized structures in the present-day universe,
containing  dark matter halos  of $ 10^{15} M_{\odot}$, thousands of galaxies, 
and  hot diffuse gas (see review by B\"ohringer \& Werner 2010).
The hot intercluster medium (ICM) constitutes the most significant baryonic component
of galaxy clusters and is observed in thermal X-ray emission with temperatures $T \sim 10^{7-8}$ K
(Voit 2005; B\"ohringer \& Werner 2010).
Many mysteries remain regarding the 
multiphase nature of the gas in the ICM and its role in 
hierarchical structure formation.
For instance, simulations and observations  suggest that  roughly 20\% of the baryonic mass budget
is expected to exist in a warm  (i.e. T $< 10^5$ K) diffuse phase (Cen \& Ostriker 1999; Dave et al. 2001; Penton et al. 2000, 2004; Lehner et al. 2007;
Danforth \& Shull 2008; Yoon et al. 2012).
This warm diffuse medium in the ICM is difficult to detect in emission and therefore absorption lines are the most direct way of probing it (Yoon et al. 2012).
In addition to the warm diffuse gas, a colder neutral component
traced by the 21-cm line is also observable.  However unlike the interstellar medium (ISM) of galaxies, 
 multiphase clouds in pressure equilibrium have not been demonstrated to exist in the ICM as of yet.

To complicate the picture of the life-cycle of gas in the ICM,
a number of authors have reported the existence of 
clumps of  neutral hydrogen emission 
(i.e. the 21-cm line) with no discernible stellar counterparts, that are seemingly not associated with other nearby galaxies (e.g. Minchin et al.
2005; Kent et al. 2007; Koopmann et al. 2008; Taylor et al. 2012, 2013; Janowiecki et al. 2015; Cannon et al. 2015).
These so-called  ``dark galaxies" have been proposed as a solution
to the ``missing satellites'' problem (Bullock 2010; Davies et al. 2006) and have a typical HI mass of 
$\sim 10^7 M_{\odot}$, which is near the detection limit of most HI surveys.

In many cases, initial detections of dark galaxies later revealed that they
either have a very small stellar component (i.e. have just begun star formation)
or are in fact tidally stripped HI streams.  An example of this latter
possibility is the source known as VIRGOHI21 (Minchin et al. 2007), a dark galaxy that 
is part of a much larger extended structure (Haynes et al. 2007). 
In some cases, these dark HI clumps are rotating, indicating that there might be a significant
amount of dark matter; however in other cases only narrow emission lines are detected,
which are not indicative of  rotational support (e.g. Taylor et al. 2012).

To prove that these gas clouds do not
have any optical counterparts and did not form via tidal
interactions is a formidable observational challenge, especially because
in many cases the objects are not spatially resolved. 
From a theoretical perspective,  SPH simulations (e.g. Villaescusa-Navarro et al. 2015) show  HI clumps with no stellar components
which exist preferentially at the periphery of galaxy clusters.  However the issue
is  complicated by the small HI masses of these objects, which for $10^7 M_{\odot}$
is at the numerical resolution limit.  Additional uncertainty about the nature of these small clumpy HI 
clouds lies in the fact that SPH simulations  tend to form more cold clumpy material than grid simulations using adaptive mesh refinement 
techniques (Agertz et al. 2007). 
Theoretical studies have argued that it will be very difficult to maintain
a star-free neutral cloud with mass greater than $10^9 M_{\odot}$ 
since the cloud will form molecular hydrogen, become Toomre unstable,  and begin to collapse (Taylor \& Webster 2005).

In this \textit{Letter} we study the origin 
 of the HI dark clumps found in the
Virgo Cluster by the ALFALFA HI Arecibo survey and in other recent surveys regarding the properties
of  ``dark" or  ``almost dark" HI clumps (Taylor et al. 2012; Taylor et al. 2013; Janowiecki et al. 2015; Cannon et al. 2015).
We investigate the possibility that HI clumps found in the Virgo cluster sample of Taylor et al. (2012) could be
held together by external pressure rather than internal gravitational support.
Pressure equilibrium of cold and warm neutral hydrogen in  the  interstellar medium of galaxies
is widely observed and noted in simulations of thermally unstable gas (e.g. Heiles \& Troland 2003; Gazol 2014); however it has yet to be
demonstrated for cold gas in the intercluster medium.
The  population of dark or nearly-dark HI clumps reported in Janowiecki et al. (2015) and  Cannon et al. (2015)
are spatially resolved or nearly resolved, and hence we can determine if they lie near the predicted
surface density threshold for H$_2$ formation.
We provide an overview of these data in Section \ref{sec:data}. We discuss the possibility that these HI clumps could be in
pressure equilibrium with the hot X-ray emitting medium and place limits on their star formation activity
in Section \ref{sec:result}.  
Finally we discuss these results in the context of future observations and simulations in Section \ref{sec:dis}, followed by our conclusions in Section \ref{sec:con}. 

\section{Data and Surveys}
\label{sec:data}

The  ALFALFA HI survey (Giovanelli et al.
2005; Haynes et al. 2011) has obtained a census of HI in more than 5,500 galaxies
over 7,000 square degrees of the Arecibo sky.  Of the extragalactic HI sources detected, fewer than 1.5\%
were not clearly identified with a stellar counterpart in the Sloan Digital Sky Survey (SDSS).
In this sample, only $\sim50$ sources were not located near another HI structure
which would suggest a tidal stripping event had occurred. Taylor et al. (2012) reported the discovery of seven dark HI sources in and around the Virgo galaxy cluster, 
with four of these sources being at the distance of Virgo. 
 The nature of these sources is unclear,
especially due to the fact that they are not resolved within the Arecibo beam at L-band (3'5).
We present the properties of the four  Taylor et al. (2012) sources with distances of 17 Mpc (in Virgo) in Table 1.
We exclude from our sample three listed sources from the catalog that are farther than 17 Mpc away and hence may not be associated with the
Virgo cluster.

\begin{table*}
\begin{center}
\caption{Taylor et al. (2012) Virgo Cluster HI Dark Clump Properties.  Bracketed values indicate errors.
\label{tab:models}}
\begin{tabular}{ccccccccc}
\hline\hline
Source & RA & Dec.    &	W50 (km/s) Ê& Log HI Mass ($M_{\odot}$)  & Dist.  to M87 (Mpc) & $P_{X-ray} \times10^{-5}$keV/cm$^{3}$  \\	
\tableline

AGESVC1 257	&  12:36:55.10(0.8)  &   +07:25:48.0(12)  	& 131(13) 	&  7.13   &	1.53	& 	1.97	\\

AGESVC1 258	& 12:38:07.20(1.0) &  +07:30:45.0(14)            & 32(18)  &	7.13	 	& 1.54   &   1.95  	\\

AGESVC1 266 &	12:36:06.50(0.9)	& +08:00:07.0(13)	& 77(21)  &      7.22	&	1.36 & 1.36	\\

AGESVC1 274 &	12:30:25.60(0.8) &	+08:38:05.0(12) &	22(4)   &	6.86	&	1.11	&	 1.11 	\\

\hline
\end{tabular}
\end{center}
\end{table*}

The sample of marginally \textit{spatially resolved} starless HI clumps to date is primarily limited to recent efforts
by the ALFALFA  (Almost) Dark Galaxy Project (Janowiecki et al. 2015; Cannon et al. 2015).
The ALFALFA (Almost) Dark Galaxy Project has been studying the
very small fraction ($\sim$0.4\%) of HI sources with no
optical counterparts and no obvious tidal connection to other sources.  
These studies attempted to find low surface brightness
optical counterparts and to spatially resolve the HI sources.  

Janowiecki et al. (2015) performed  deep optical observations of systems 
identified by the ALFALFA HI survey with the WIYN telescope (to detect possible low surface brightness
stellar systems) and HI Synthesis map with WSRT to spatially resolve the systems 
and obtain their HI masses. 
Cannon et al. (2015) followed up several ALFALFA sources with VLA observations and compared them to archival GALEX and SDSS observations in order
to constrain star formation rates and sizes of the  clumps.
We present properties of the  ``Dark" and  ``Almost Dark" HI sources
from Janowiecki et al. (2015) and Cannon et al. (2015) in Table 2.
Most of these sources are not associated with the Virgo cluster but 
provide a spatially resolved sample with optical followup to compare with the Taylor et al. (2012) dark clumps.
 We note that most of the Cannon et al. (2015) sources are only marginally resolved in the HI beam and therefore the values
reported in Table 2 represent upper limits on the projected size of the clumps.

\begin{table*}
\begin{center}
\caption{ Properties of the ALFALFA
(Almost) Dark Galaxy Project from Cannon et al. 2015 and Janowiecki et al. 2015
\label{tab:models}}
\begin{tabular}{cccccc}
\hline\hline
Source &  Log HI Mass ($M_{\odot}$) &  HI size (kpc) & SFR ($M_{\odot} yr^{-1}$)& Stellar Mass ($M_{\odot}$) &W50 (kms$^{-1}$)  \\	
\hline \\
Janowiecki et al. (2015) & & & & & \\
\tableline

\hline \\

AGC229383 & 8.08 & 12x4 &N/A & $<3.7\times10^{5}$&  34(1)\\
AGC229384 & 8.30 & 14x10 & N/A& $<3.4\times10^{5}$&  27(1)\\
AGC229385 & 8.86 & 28x16 & $\approx 4.1-6.9\times 10^{-3}$& 1.5$\times10^{6}$& 59(8) \\
\hline \\
Cannon et al. (2015) & & & &\\
\tableline

\hline \\
AGC193953 & 8.21 & 9.4& N/A &2.2$\times10^{6}$& 32  \\
AGC208399 & 7.41 & 2.5 & 5.1$\times10^{-5}$&1.7$\times10^{6}$ &31  \\
AGC208602 & 8.07 & 4.4 & $<2.6\times10^{-4}$& $<2.5\times10^{6}$&  50  \\
AGC226178 & 7.60 & 3.9  &0.0017& 1.3$\times10^{6}$& 28   \\
AGC233638 & 9.51 & 32.6 & 0.047& 2.2$\times10^{8}$& 62 \\
\hline\hline
\end{tabular}
\end{center}
\end{table*}


\section{External Pressure Support in Virgo and Low Star Formation Rates}
\label{sec:result}

The sources reported in Taylor et al. (2012)  are particularly  
intriguing as Virgo is the nearest galaxy cluster and can be studied with high spatial resolution.
Observations of the Virgo cluster have mapped
 X-rays (e.g.,
B\"ohringer et al.  1994; Urban et al. 2011), atomic hydrogen
(Giovanelli et al. 2007; Popping \& Braun
2011; Yoon et al. 2012),  dust (Davies et al. 2010; Baes et al. 2010), and stellar distributions (Co\^te et al.  2004; Mihos et al. 2005).

\begin{figure*} \begin{center}
\includegraphics[width=0.79\textwidth]{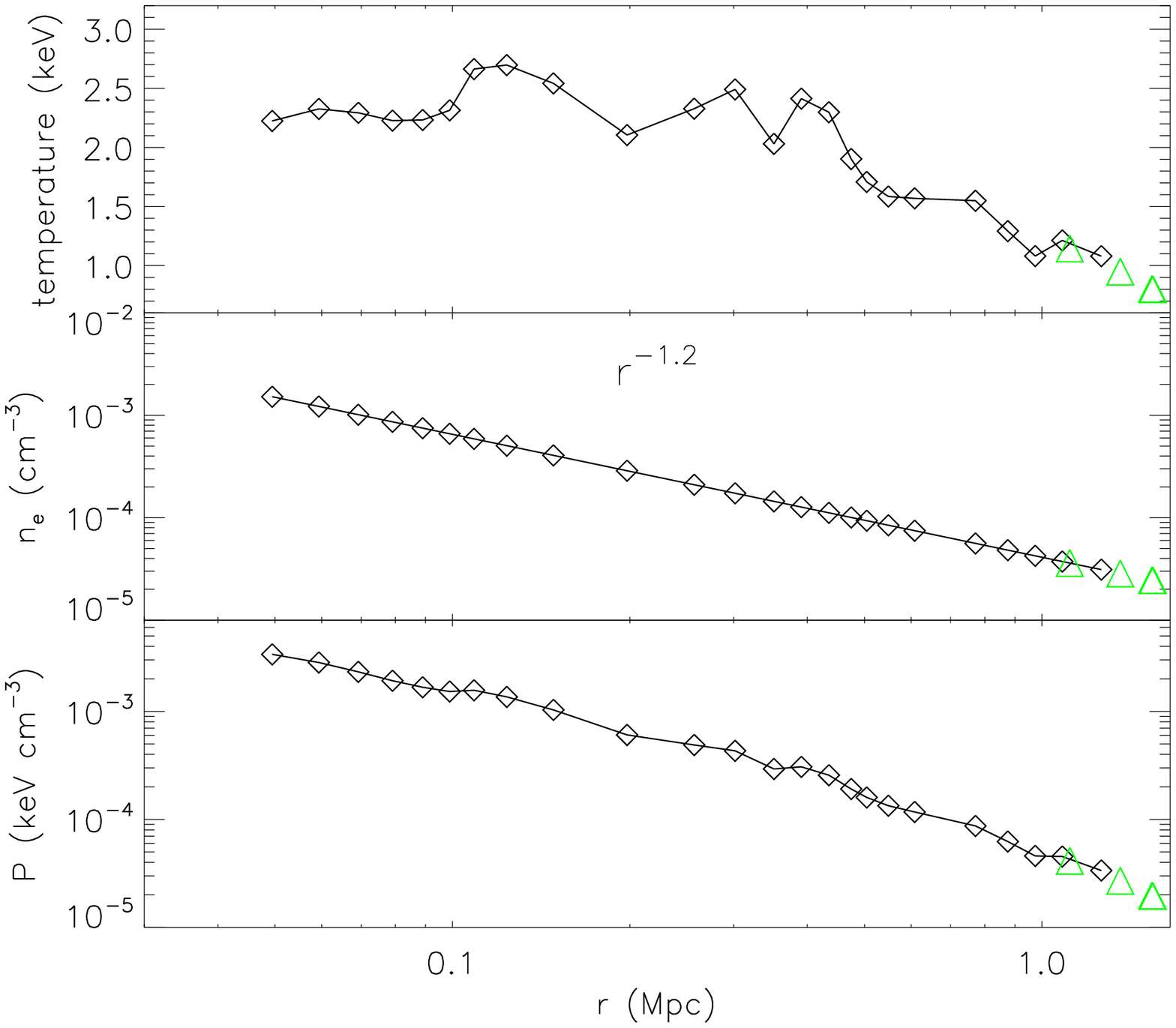}

\caption[ ]{X-ray properties in the Virgo Cluster: The deprojected temperature (top), electron density ($n_e$, center), pressure
$(P_{X-ray}$, bottom) profiles are adopted from the Urban et al. (2011) XMM-Newton data set. 
The location of the four Arecibo HI dark clumps of Taylor et al. (2012) are linearly extrapolated from the Urban et al. (2011) profiles and then  overplotted as green
triangles.  We note that source AGESVC1-257 and AGESVC1-258 have identical distances from M87 (see Table 1) and hence overlap in the plot.}
\label{fig:xray} \end{center} \end{figure*}

Although the four Taylor et al. (2012) sources in Virgo are not resolved in the ALFALFA beam,
we can estimate what their sizes would be if the HI  clumps were in pressure equilibrium 
with the local surrounding hot X-ray emitting medium  in Virgo.
Urban et al. (2011) performed thirteen XMM-Newton pointings covering
the Virgo Cluster from the center out to a radius  just beyond the Virial radius at 1.2 Mpc. 
We list the distances of the Taylor et al. (2012) sources from the center of Virgo (M87) in Table 1, which can then be
directly related to the X-ray pressure profile vs. distance in Urban et al. (2011).
We adopt the Urban et al. (2011) measured density profile best-fit power-law model to describe the ICM
electron density profile as $n_e \propto r^{-1.2}$ and plot the temperature, electron density, and the pressure profiles in 
Figure \ref{fig:xray}. The locations of the four Arecibo HI dark clumps of Taylor et al. (2012) are linearly extrapolated from the 
Urban et al. (2011) profiles and overplotted as green
triangles. From these points we list the values of the X-ray pressure ($P_{X-ray}$) at the clump locations
in Table 1.

Given the X-ray pressure at each location, the measured HI mass and the measured value of the
line width (listed in Table 1), we can estimate the required size of the clumps for them  to be in 
pressure equilibrium with the surrounding hot X-ray gas, i.e. for $P_{X-ray}=P_{HI}$,
\begin{equation}
P_{HI}=\frac{\rho k_bT_{eff}}{\mu m_p}
\end{equation}
where $\rho=\mu m_p n$ is the gas mass density, $\mu$ is the mean molecular weight, $k_b$ is the Boltzmann constant,  and $m_p$ is the proton mass.
$T_{eff}$ is measured from spectral line widths for the ALFALFA dark sources,
 which measure the effective total temperature given by thermal and turbulent pressure support:

\begin{equation}
\frac{1}{2}\sigma_{tot}^2=\frac{kT_{eff}}{\mu m_p}
\end{equation}

with
$P_{HI}=\frac{1}{2} \rho_{HI}\sigma_{tot}^2$
where
$\rho_{HI}=\frac{M_{HI}}{\frac{4}{3}\pi r^3}$
 where $M_{HI}$ is the HI mass of the cloud, $r$ is the physical size of the HI dark clump, which is unknown in the case of the Taylor et al. (2012) sample
since these clouds are unresolved in the Arecibo beam.  We therefore get, 
\begin{equation}
P_{HI}=P_{X-ray} \sim \frac{1}{2}\sigma_{tot}^2 \frac{M_{HI}}{\frac{4}{3}\pi r^3}.
\end{equation}

Given the HI mass, velocity dispersion, and value of $P_{X-ray}$, we estimate the sizes of the clumps in
pressure equilibrium with the ICM.  Figure 2 plots the predicted sizes vs. HI masses of as purple triangles, 
which lie between 1 kpc and 10 kpc,  below the Arecibo beam size at the distance of Virgo.
We overplot theoretical lines of constant HI temperature (shown in the legend) assuming pressure equilibrium with
$P_{X-ray}=10^{-6}$ keV/cm$^{3}$.

\begin{figure*} \begin{center}
\includegraphics[width=0.69\textwidth]{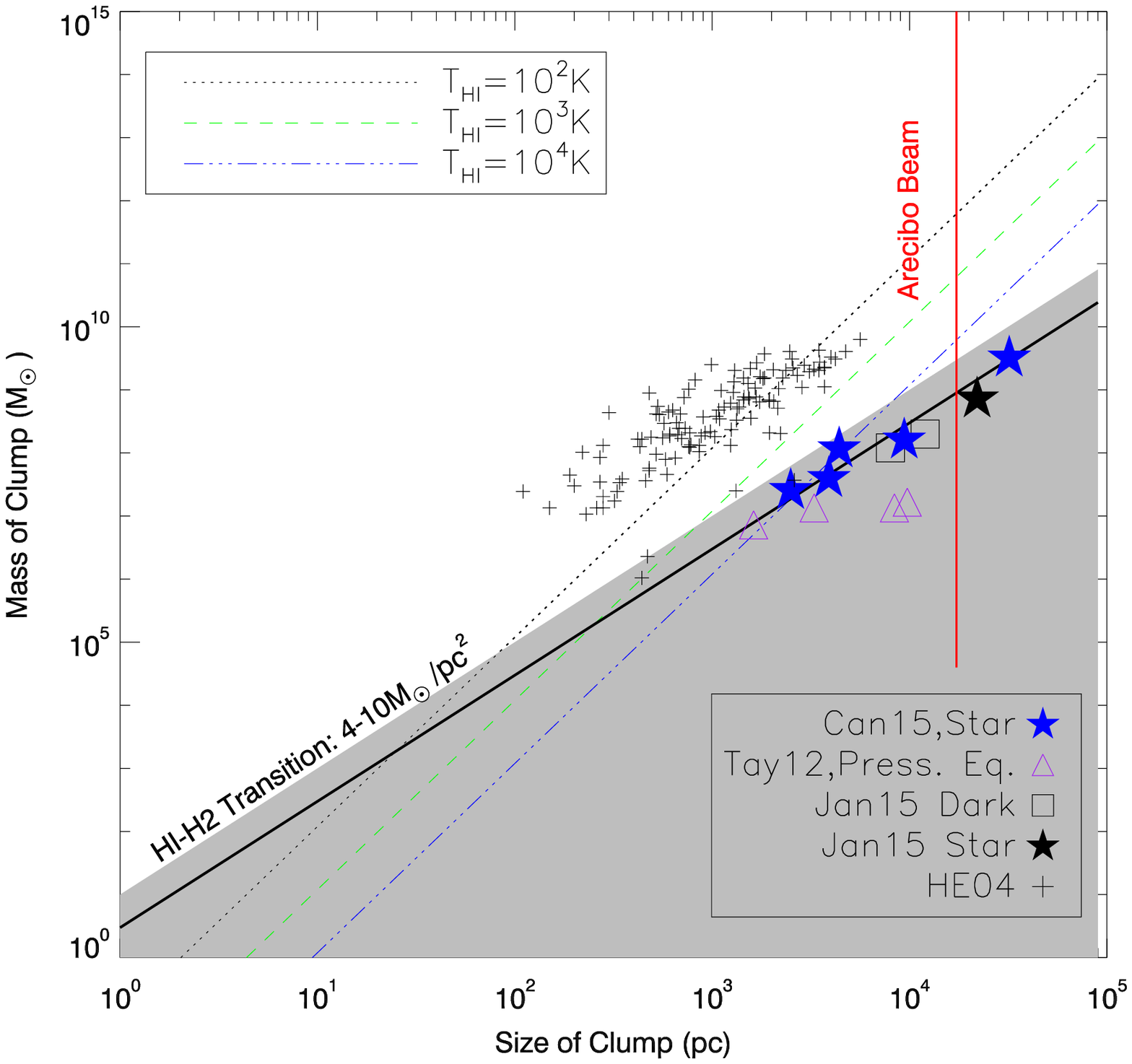}

\caption[ ]{Size vs. mass relationship for starless or nearly starless HI clumps. The present work computes the sizes of the Taylor et al. (2012) sources (denoted by purple triangles) by assuming pressure equilibrium with the surrounding X-ray gas.  The mass-size relation for these objects is consistent with that expected from typical HI temperatures ($10^2-10^4$ K) for these objects and the assumption of pressure equilibrium (taking $P_{X-ray}=10^{-6}$ keV/cm$^{3}$), shown by the three dashed lines. We also show the average measured sizes and masses of similar sources from Cannon et al. (2015) and Janowiecki et al. (2015) that are resolved or nearly resolved; they are consistent with the sizes we predict here for the Taylor et al. (2012) sources. The shaded region roughly corresponds to objects unable to form stars since they are below the critical surface density for ${\rm H}_2$ formation. Sources with detected stellar components are denoted by star symbols while those without detected stellar counter parts are denoted by squares; most of the sources with detected stellar components are only marginally able to form stars. Finally, we show size vs. mass for  dwarf galaxies from Hunter \& Elmegreen (2004, small cross symbols), which are roughly comparable in size and mass to our ALFALFA starless HI clump sample. 
}
\label{fig:sizemass} \end{center} \end{figure*}

We now determine if these estimated sizes are consistent with the fact that the Taylor et al. (2012) HI sources seemingly are starless.
The primary ways for supporting these clouds against collapse and star formation are through 
turbulence, thermal pressure,  and rotation (not observed in the Taylor et al. 2012 linewidth profiles).
In general, low mass protogalactic disks will become Toomre unstable if they are able to cool 
efficiently via molecular line cooling. H$_2$ formation can be  limited by photodissociation from the X-rays emitted by the cluster
and therefore a critical shielding surface density of HI  is required.  The
thickness of the shielding layer can be characterized by a critical
surface density of gas, $\Sigma_{crit}$, below which the H$_2$ is photodissociated and star formation will not occur.

A number of studies have been devoted to estimating the $\Sigma_{crit}$ value for the HI to H$_2$ transition both in the context of our
own Milky Way Galaxy (Federman, Glassgold \& Kwan 1979; Lee et al. 2012, Sternberg et al. 2014; Burkhart et al. 2015) as well as external galaxies (Elmegreen 1993; Krumholtz, McKee \& Tumlinson 2008).  
Generally these studies find values between 4-10  M$_{\odot}$pc$^{-2}$.
Taylor \& Webster (2005) estimate $\Sigma_{crit}=4$ M$_{\odot}$pc$^{-2}$
specifically for conditions pertaining to dark galaxies embedded in a hot UV/X-ray
medium.  However, pressure confined clouds will transition at a lower surface density (Schaye 2001), which depends on the volume density of the cloud.
 In Figure 2 we overplot the  Tayler \& Webster (2005) value for the HI-H$_2$ transition, 
$\Sigma_{crit}=4$ M$_{\odot}$pc$^{-2}$, as a black straight line.  We shade in grey the surface density values below $\Sigma=10M_{\odot}pc^{-2}$,
below which we expect H$_2$ formation is suppressed.
As expected from the theoretical considerations of Taylor \& Webster (2005) 
all the purple triangles lie below the HI-H$_2$ transition and may not be star-forming as long as they have sizes given by the assumption of  pressure equilibrium.
The Arecibo beam size and and the HI-H$_2$ transition provide upper and lower bounds on  the sizes of these sources and in all four cases,
the assumption of pressure equilibrium with the X-ray gas yields sizes that satisfy these bounds.

For consistency we overplot the average sizes and masses of sources from the studies by Cannon et al. (2015) and  Janowiecki et al. (2015), most of which are resolved or nearly resolved.
Sources that have detected stellar components are denoted with star symbols while those without detected
stellar counterparts are denoted with squares.  In all cases these nearly starless or starless HI clumps
have surface densities which closely bracket the critical surface density star formation threshold predicted in Taylor \& Webster (2005).
This suggests that objects with this surface density are on the cusp of star formation. Since the HI signal is proportional to $M_{HI}$,
there is a bias towards detecting HI clumps near the upper envelope of the shaded region.
For comparison, Figure 2 also shows the size vs. HI masses for a sample of dwarf galaxies reported in Hunter \& Elmegreen (2004, small crosses) which have similar mass and size ranges
as the ALFALFA starless HI clump sample.  Just as  expected for star-forming dwarf galaxies, all of these points lie above the critical surface density.

\section{Discussion}
\label{sec:dis}

Under the assumption of pressure equilibrium,  two of the Taylor et al. (2012) objects (sources 266 and 257) are significantly below the critical surface density implying that H$_2$ formation
is suppressed in them.  In addition, the HI clumps discussed here have turbulence plus thermal Jeans masses, 
\begin{equation}
M_J=\frac{3k_bT_{eff}r}{2Gm_p}
\end{equation}
which are all roughly an order of magnitude above the measured HI mass and therefore should be on average stable against collapse.
The Taylor et al. (2012) sizes we predict assuming pressure equilibrium with the X-ray ICM in Virgo are similar
to other nearly starless HI sources in other ALFALFA survey studies (Cannon et al. 2015; Janowiecki et al. 2015).  Other properties
of the HI clumps in these studies, such as their HI masses and velocity dispersions, are also similar (see Tables 1 and 2).
Since all of these sources populate the same parameter space in terms of low or no star formation, masses, narrow velocity dispersions, and sizes, 
their origin and survival in the ICM may be similar.  The sizes estimated in the present work are consistent with other studies that show giant HI clumps
can exist in pressure equilibrium on kpc scales within host galaxies (Behrendt, Burkert, \& Schartmann 2016).

What would the lifetime of HI dark clouds in pressure equilibrium be against thermal evaporation?
Cowie and McKee(1977) have investigated the case of classical thermal evaporation.  The time scale for evaporation of the interstellar HI gas in the host ICM depends on 
the size of the cloud and ICM temperature (Veilleux et al. 1999):
\begin{equation}
t_{cond}=10^9 n_c R^2_{kpc} \Big(\frac{T_{ICM}}{10^7 \textrm{K}} \Big)^{-5/2}  \Big( \frac{\ln \Lambda}{30}\Big) \textrm{yr}
\end{equation}
where $n_c$ is the HI cloud density in cm$^{-3}$, $R_{kpc}$  the cloud radius in kpc, T$_{ICM}$ is the temperature of
the surrounding hot ICM, and $\ln \Lambda$ is the Coulomb logarithm. 
For ISM HI clouds of typical sizes (e.g. 10 pc) and densities (n$_c \sim1$ cm$^{-3}$) embedded in a T$_{ICM}=10^7$ K gas,  the evaporation timescale 
is short at $\sim10^5$ years.  However, the HI clouds in our sample are large (i.e. the size of dwarf galaxies) and therefore
have evaporation timescales greater than $10^9$ years for the same HI density and ICM temperature. 
Additionally, cold fronts are preserved in the ICM, presumably due to magnetic insulation (see Hummel \& Saikia 1991; Markevitch \& Vikhlinin 2007), 
which could also prevent evaporation.
Thermal conduction can also quickly heat the gas and suppress star formation.
Simulations have shown that heat conduction can actually help cold clouds survive passage through hot gas by slowing
the dissipative effects of the Kelvin-Helmholtz instability (Vieser \& Hensler 2007).  
However,  If these clouds are on typical orbits within the cluster than pressure fluctuations and tidal shear may increase the
surface density by more than an order of magnitude and trigger star formation.  Future modeling of dark HI clouds in clusters
should be performed in order to determine if star formation may be triggered due to external environmental factors.

Numerical simulations  can lead to better modeling of starless HI clouds in galaxy clusters.
Recent SPH simulations have shown that  a significant fraction of HI in galaxy clusters
actually resides in small  clouds in the ICM rather than as HI mass within galaxies themselves (Villaescusa-Navarro et al. 2015).
More neutral hydrogen  clouds are also found at the outskirts of the cluster rather than towards the center,
in agreement with the locations of the Taylor et al. (2012) dark galaxies.

Our study suggests that higher resolution follow-up observations, as well as higher resolution simulations, will be  key to understanding the nature of the
starless HI clumps in galaxy clusters.  
The HI mass of these sources range between $10^7-10^9 M_{\odot}$ which is near the resolution limit of
both the ALFALFA survey and numerical simulations.  Therefore the currently detected sources, while seemingly rare at these mass ranges, may represent the tip of the iceberg
in terms of actual population of HI clouds in pressure equilibrium with the ICM.
Objects that have been followed up by the  ALFALFA  (Almost) Dark Galaxy Project 
are either not in the Virgo cluster (i.e. the Cannon et al. 2015 sample) or possibly associated with Virgo but at uncertain distances (e.g. $D=25$ Mpc in the case of sources of Janowiecki et al. 2015).
Optical follow-up of these objects is key to ascertain if they have low surface brightness stellar populations or not.  Our study
suggests that the total HI mass is not the main driver for the formation of stars. Rather, the HI surface density
will ultimately  determine if we should expect to observe stellar populations  in these objects. 
If there is significant thermal conduction between the hot ICM and cold HI gas we also expect the existence of
intermediate temperature gas tracers to be present and observed in absorption, i.e. Lyman-$\alpha$, the CIV  emission line or HeII   
(Yoon et al. 2012; Sparks et al. 2012). Follow-up observations of these tracers in and around dark galaxies would be of considerable interest.

\section{Conclusion}
\label{sec:con}

Using data from the Taylor et al. (2012) ALFALFA Arecibo HI survey of galaxies and the Virgo cluster X-ray pressure profile from XMM-Newton,
 we investigated the possibility that starless dark HI clumps, also known as  ``dark galaxies",  could be held together
by support from the surrounding hot X-ray emitting ICM.
We predict the sizes of these sources assuming pressure equilibrium with the ICM and find they range from 1 kpc  to 10 kpc.  These scales suggest surfaces densities just below the critical threshold for the HI-H$_2$ transition, 
consistent with the fact that no stars are observed in the clumps. 
Based on the predicted HI surface density of the Virgo sources, as well as a sample of other resolved ALFALFA HI dark clumps with follow-up optical/radio observations,
we predict that most of the HI dark clumps are on the cusp of forming stars.
Follow-up observations of the Taylor et al. (2012) Virgo sources are needed in order to spatially resolve the HI
distribution and determine if the sources are near the sizes predicted by our pressure equilibrium arguments.  Numerical
simulations of the ICM should focus on resolving the masses and sizes found here.
If confirmed, this would be the first detection of pressure-confined HI clumps in the multiphase ICM.

\acknowledgments
B.B. acknowledges support from the NASA Einstein Postdoctoral Fellowship. A. L. was supported in part by NSF grant AST-1312034.
The authors thank the anonymous referee,  Prof. Andreas Burkert, Prof. Joop Schaye, Dr. Zachary Slepian, and Prof. Rhys Taylor for insightful comments.

\end{document}